# Systematic Theoretical Study on the Interstellar Carbon Chain Molecules

Emmanuel E. Etim[1,2], Prasanta Gorai[3], Ankan Das[3*], Sandip K. Chakrabarti[4,2], Elangannan Arunan[1]

[1] Inorganic and Physical Chemistry Department, Indian Institute of Science Bangalore, India-560012
[2] Department of Chemical Sciences, Federal University Wukari, Katsina-Ala Road, P.M.B. 1020 Wukari, Taraba State, Nigeria
[3] Indian Centre for Space Physics, 43 Chalantika, Garia Station Road, Kolkata 700 084, India
[4] S. N. Bose National Centre for Basic Sciences, Salt Lake, Kolkata 700098, India
*email: ankan.das@gmail.com

**Abstract**: In an effort to further our interest in understanding basic chemistry of interstellar molecules, we carry out here an extensive investigation of the stabilities of interstellar carbon chains; $C_n$, $H_2C_n$, $HC_nN$ and $C_nX$ (X=N, O, Si, S, H, P, H$^-$, N$^-$). These sets of molecules accounts for about 20% of all the known interstellar and circumstellar molecules, their high abundances therefore demand a serious attention. High level *ab initio* quantum chemical calculations are employed to accurately estimate enthalpy of formation, chemical reactivity indices; global hardness and softness; and other chemical parameters of these molecules. Chemical modeling of the abundances of these molecular species has also been performed. Of the 89 molecules considered from these groups, 47 have been astronomically observed, these observed molecules are found to be more stable with respect to other members of the group. Of the 47 observed molecules, 60% are odd number carbon chains. Interstellar chemistry is not actually driven by the thermodynamics, it is primarily dependent on various kinetic parameters. However, we found that the detectability of the odd numbered carbon chains could be correlated due to the fact that they are more stable than the corresponding even numbered carbon chains. Based on this aspect**,** the next possible carbon chain molecule for astronomical observation in each group is proposed. The effect of kinetics in the formation of some of these carbon chain molecules is also discussed.

## 1.0 Introduction

As more and more interstellar and circumstellar molecules are observed, our understanding of the chemistry of these molecules is either strengthened or challenged depending on whether the new molecules fall within or outside the boundaries of their expected basic chemistry. These observations also put a constraint on the different chemical models. The fields of astronomy, astrophysics, astrochemistry and astrobiology have witnessed dramatic changes in the last few decades as a result of the advances in science and technology. In addressing the chemical origin of life and for the proper understanding of the solar system, there has been a constant and consistent searching for molecules in the interstellar and circumstellar media resulting in unique observation of about 200 different molecular species till date. The year 2014 has seen unparalleled observations of nine new interstellar and circumstellar molecules including the first branched chain molecule in space, namely, isopropyl cyanide, an important biological molecule; urea, the sulphur analogue (ethyl mercaptan) of the well known interstellar alcohol; ethanol, among others and the confirmation of many others (Remijan et al. 2014; Belloche et al. 2014; Agúndez et al. 2014; Anderson & Ziurys 2014; Cernicharo 2014; Kolesniková et al. 2014).

Almost on a regular basis, carbon chain molecules such as $C_n$, $H_2C_n$, $HC_nN$ and $C_nX$ (X=N, O, Si, S, H, P, H$^-$, N$^-$) are astronomically observed. Till date, the largest claimed straight chain interstellar molecule is the cyanopolyne, $HC_{11}N$; a linear carbon chain molecule. Among the carbon chains, there seems to be no easily understood trend or pattern within the observed molecules. In the $C_nN$ and $C_nN^-$ chains, with the exception of $C_2N$, only the odd number carbon chains are observed from n=1 to n=6 (Jefferts et al. 1970; Friberg et al. 1980; Guélin et al. 1998; Thaddeus et al. 2008, Agúndez et al. 2010; Cernicharo et al. 2008; Anderson & Ziurys 2014). Whereas in the $C_nH$ chains, both odd and even number chains are observed between 1 and 8 but in the $C_nH^-$, only even number chains are observed (Swings & Rosenfeld, 1937; Tucker et al., 1974; Guelin et al., 1978; Thaddeus et al., 1985a; Cernicharo et al., 1986, 2007; Suzuki et al., 1986; Cernicharo & Guelin, 1996; Guelin et al., 1997; McCarthy et al., 2006; Brunken et al., 2007). Among the cyanopolyne chains, between 6 and 12 carbon atoms, only odd number chains are observed (Snyder & Buhl 1971; Turner, 1971; Avery et al., 1976; Kroto et al., 1978; Broten et al., 1978; Guelin & Cernicharo, 1991; Bell et al., 1997; Cernicharo et al., 2004) while in the $C_nSi$ group, all the molecules between carbon 1 and 4 have been observed (Thaddeus et al.1984; Cernicharo et al. 1989; Ohishi et al. 1989; Apponi et al. 1999). This type of confusing trend is observed among the different groups of linear interstellar carbon chains.

As part of our interest in understanding the basic chemistry of interstellar and circumstellar molecules, the present work focuses on the interstellar and circumstellar carbon chains; $C_n$, $H_2C_n$, $HC_nN$ and $C_nX$ (X = N, O, Si, S, H, P, H$^-$, N$^-$). High level quantum chemical calculations are applied to accurately determine the enthalpies of formation, chemical reactivity indices; global hardness and softness; and other chemical parameters of these molecules. Chemical modeling of the abundances of these molecular species has also been performed; the results are used in accounting for the general observation of more odd carbon chains than the even carbon chains and the dominance of the even carbon chains in a few cases. The possible candidates for astronomical observation among the carbon chain molecules are also considered.

**2.0 Computational Details**: The results of the quantum chemical calculations reported in this work were obtained using Gaussian 09 suite of programs (Frisch et al., 2009). In order to estimate accurate enthalpies of formation that are in good agreement with the experimental values (where available) for the different carbon chains considered in this work, the G4 composite method was employed. The choice of the G4 composite method is borne out of experience as it has been shown to compute enthalpies of formation to chemical accuracy (Curtiss et al. 1998, 2007a,b). The geometries of these molecules were optimized at the G4 level, harmonic vibrational frequency calculations were used to characterize the stationary nature of all the structures with equilibrium species possessing only real frequencies. The reported zero-point corrected standard enthalpies of formation of the carbon chains were calculated from the optimized geometries. The enthalpies of formations were calculated from atomization energies. With G4 computational method and accurate experimental values of standard enthalpy of formation of the constituent elements involved, very high accurate enthalpies of formation can be estimated for different set of molecular systems. Atomization energies (sometimes referred to as the total dissociation energies) were evaluated using the calculated values of energies (sum of electronic and zero-point energy corrections). In calculating the enthalpy of formation ($\Delta_f H$) at 0 K for all the molecules reported in this study, the experimental values of standard enthalpy of formation of elements C, H, O, N, P, Si and S reported in literature were used (Ochterski, 2000). Chemical reactivity indices such as global hardness ($\eta$) and softness (S) have been calculated for different carbon chains examined in this study. For the computation of S and $\eta$, we use DFT method with B3LYP/6-311g++(d,p) basis set of Gaussian 09 program. First, we calculate the Ionization potential (I) of the desired species by computing the energy (ground state optimized energy) difference between the cation and neutral and Electron affinity (A) by computing the energy difference between the neutral and anion. Chemical hardness is calculated by $\eta=(I-A)$ (Parr & Pearson 1983) and softness is calculated by $1/\eta$. The chemical modelling of the abundances of these carbon chain species is described alongside with the obtained results under the discussion section.

**3.0 Results and Discussion**: The results of the quantum chemical calculations at the G4 level of theory obtained for all the carbon chain molecules are presented and discussed in this section (3.1 to 3.11). In all the cases, the reported zero-point corrected standard enthalpies of formation ($\Delta_f H^O$) are in kcal/mol. Section 3.12 examines the effect of kinetics on the formation processes of these carbon chains. In section 3.13, chemical reactivity indices and chemical modelling supporting the observed results in the previous sections are presented and discussed before the final conclusion.

**3.1 $C_n$ Chains**: Apart from their importance in astrophysics, astrochemistry, and astrobiology, pure carbon chain molecules such as $C_2$, $C_3$, and $C_5$ are of great interest in combustion chemistry, flames and propellants. Table 1 shows enthalpy of formation and current astronomical status of carbon chain ($C_n$) molecules considered in this study. Rotational spectroscopy has been the dominant tool in the chemical examination of interstellar and circumstellar media with about 90% of all known interstellar and circumstellar molecules observed via their rotational transitions. However, the application of rotational spectroscopy in this field is hampered by the fact that it is only limited to molecules with permanent dipole moment (Etim and Arunan, 2015). The $C_n$ chain molecules have no permanent dipole moment; hence their pure rotational transitions cannot be measured. All the known interstellar $C_n$ chain molecules have been observed in the infrared region through their vibrational transitions.

From Table 1, there is a steady increase in the enthalpy of formation of the $C_n$ chains as the chain length increases. With the exception of $C_4$ which is yet to be observed. All the lower members of the chains with lower enthalpies of formation have all been detected (Souza & Lutz, 1977; Hinkle et al., 1988; Cernicharo et al. 2000; Bernath et al., 1989). With respect to the effect of thermodynamics, the lower members of the chains with lower enthalpies of formation are more stable, than the higher members of the chains with higher enthalpies of formation. Figure 1 shows the plot of the enthalpy of formation of the $C_n$ chain molecules estimated at the G4 level. The lower members of the chains remain the most likely candidates for astronomical observation as compared to the higher members of the chains. As $C_4$ is yet to be observed and $C_5$ has already been observed, it is likely that $C_7$ having a larger enthalpy of formation may be observed before $C_6$ having a lower enthalpy of formation.

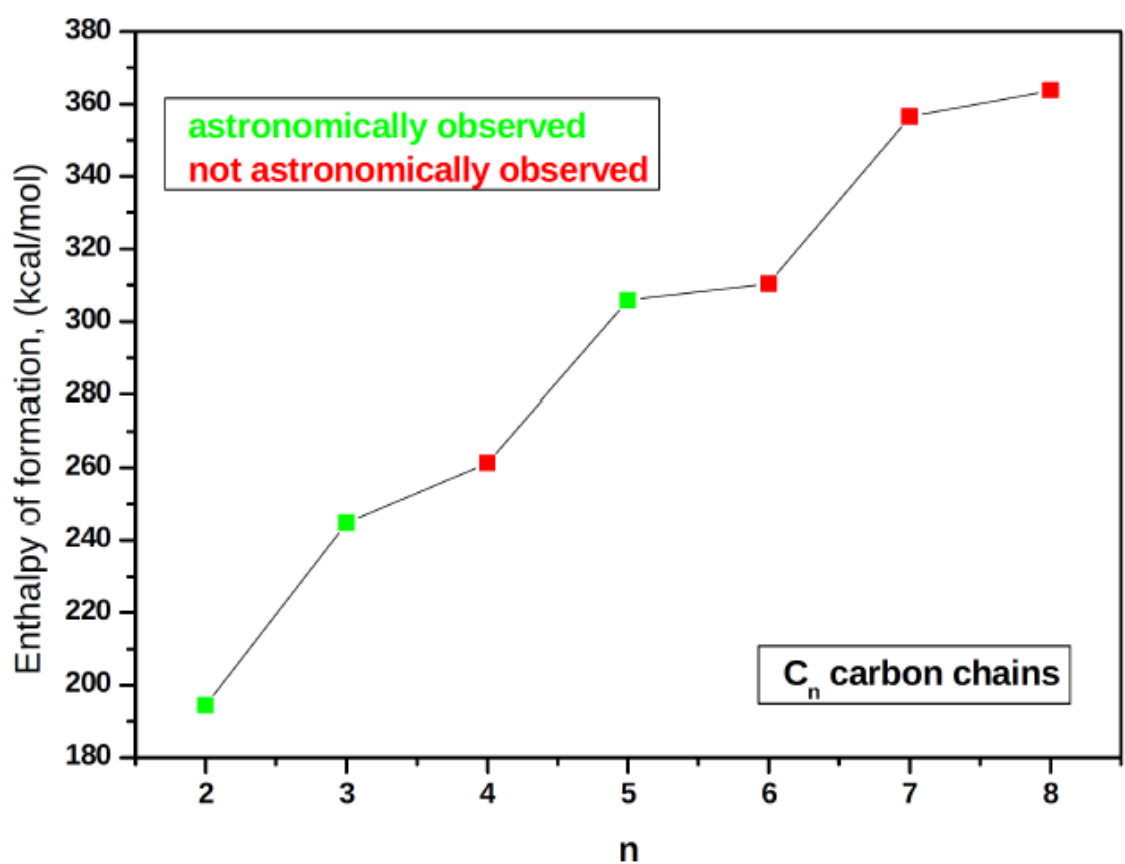

**Figure 1:** Plot showing the $\Delta_f H^O$ for $C_n$ chain molecules

**Table 1:** $\Delta_f H^O$ for $C_n$ chains and current astronomical status

| n | $\Delta_f H^O$ (kcal/mol) | Astronomical status |
|---|---|---|
| 2 | 194.432 | [a]Observed |
| 3 | 244.785 | [b]Observed |
| 4 | 261.328 | Not observed |
| 5 | 305.822 | [c]Observed |
| 6 | 310.379 | Not observed |
| 7 | 356.501 | Not observed |
| 8 | 363.736 | Not observed |

[a]Souza & Lutz,1977; [b]Hinkle et al., 1988; Cernicharo et al. 2000; [c]Bernath et al., 1989.

**3.2 $C_nO$ Chains**: The $C_nO$ group presents an interesting set of carbon chain molecules containing two (carbon and oxygen) of the four most important biological elements. Unlike the $C_n$ chains that lack permanent dipole moment, the $C_nO$ chains all possess permanent dipole moment, hence, all the known interstellar/circumstellar $C_nO$ molecules have been observed via their rotational transition spectra. The enthalpy of formation and current astronomical status for all the $C_nO$ molecules considered in this study are presented in Table 2 while Fig. 2 depicts the plot of the enthalpy of formation for these molecules. Of the six molecules considered in the $C_nO$ group, the first three (n=1 to 3) have been astronomically observed (Smith & Stecher, 1971; Ohishi et al. 1991; Matthews et al. 1984). The trend of the enthalpy of formation is the same as in the case of the $C_n$ chain molecules. The odd number carbon chain molecules are found to be more stable (lower enthalpy of formation value) that their corresponding even number carbon chain molecules. The interstellar chemistry of oxygen and sulphur is well coupled, with over 80% of all the known S-containing interstellar molecules having their corresponding O-containing molecules as known interstellar molecules. Also, in most of the cases, the abundance of S-compound relative to its O-analogue is approximately equal to the cosmic S/O ratio (Linke et al. 1979; Frerking et al. 1979). The high probability of $C_5O$ as an interstellar molecule stems from two important facts; $C_5O$ has lower enthalpy of formation as compared with $C_4O$ and $C_6O$, it is therefore more stable in ISM, thus can possibly be observed as compared to $C_4O$ and $C_6O$. Secondly, $C_5S$, the sulphur analogue of $C_5O$ has recently been observed (Agúndez et al. 2014) and O-containing molecules are known to be more abundant in ISM than their corresponding S-analogue. Hence astronomical observation of $C_5S$ is a clear indication of the presence of $C_5O$ in detectable abundance in ISM.

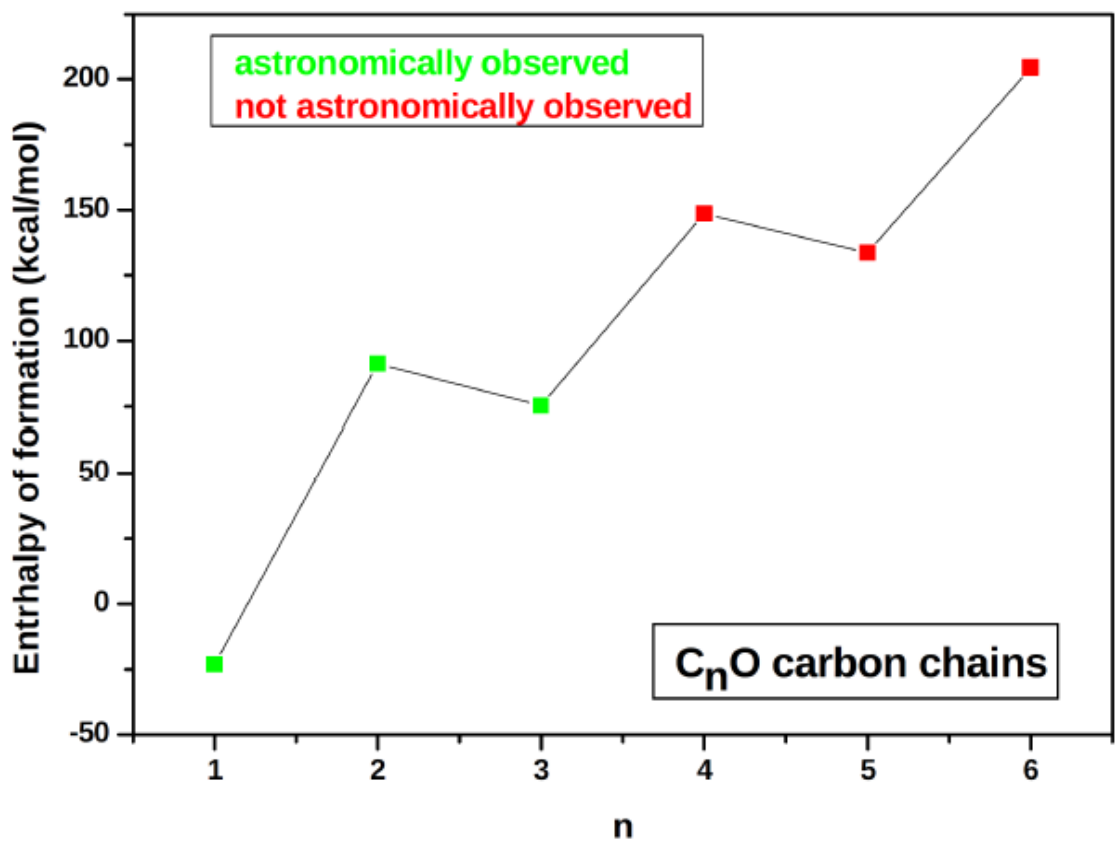

**Figure 2:** Plot showing the $\Delta_f H^O$ for $C_nO$ chain molecules

**Table 2:** $\Delta_f H^O$ for $C_nO$ chains and current astronomical status

| n | $\Delta_f H^O$ (kcal/mol) | Astronomical status |
|---|---|---|
| 1 | -23.127 | [a]Observed |
| 2 | 91.369 | [b]Observed |
| 3 | 75.328 | [c]Observed |
| 4 | 148.640 | Not observed |
| 5 | 133.782 | Not observed |
| 6 | 204.526 | Not observed |

[a]Smith & Stecher, 1971; [b]Ohishi et al. 1991; [c]Matthews et al. 1984

**3.3 $C_nS$ Chains:** IRC+10216, the carbon-star envelope has emerged as one of the richest sources of molecules with over eighty different molecular species so far observed in it. Alongside other molecular clouds, all the known $C_nS$ molecules were observed in IRC+10216. As discussed in the previous section, the interstellar chemistry of oxygen and sulphur is very similar in many ways. Of the four ($CS$, $C_2S$, $C_3S$, $C_5S$) observed molecules in the $C_nS$ group (Penzias et al. 1971; Saito et al. 1987; Yamamoto et al.1987; Cernicharo et al. 1987; Agúndez et al. 2014), 3 ($CS$, $C_2S$, $C_3S$) three have the corresponding O-analogues as known interstellar molecules. Whereas $C_5S$ has been astronomically observed, $C_4S$ and $C_6S$ are not. This could be correlated with its enthalpy of formation value (Table 3) as compared to those of $C_4S$ and $C_6S$. As shown in Table 3 and Figure 3, all the odd number $C_nS$ chains have lower enthalpy of formation than their corresponding even number $C_nS$ chains, hence they are more stable, likely more abundant and can thus be more easily observed than the even number counterparts.

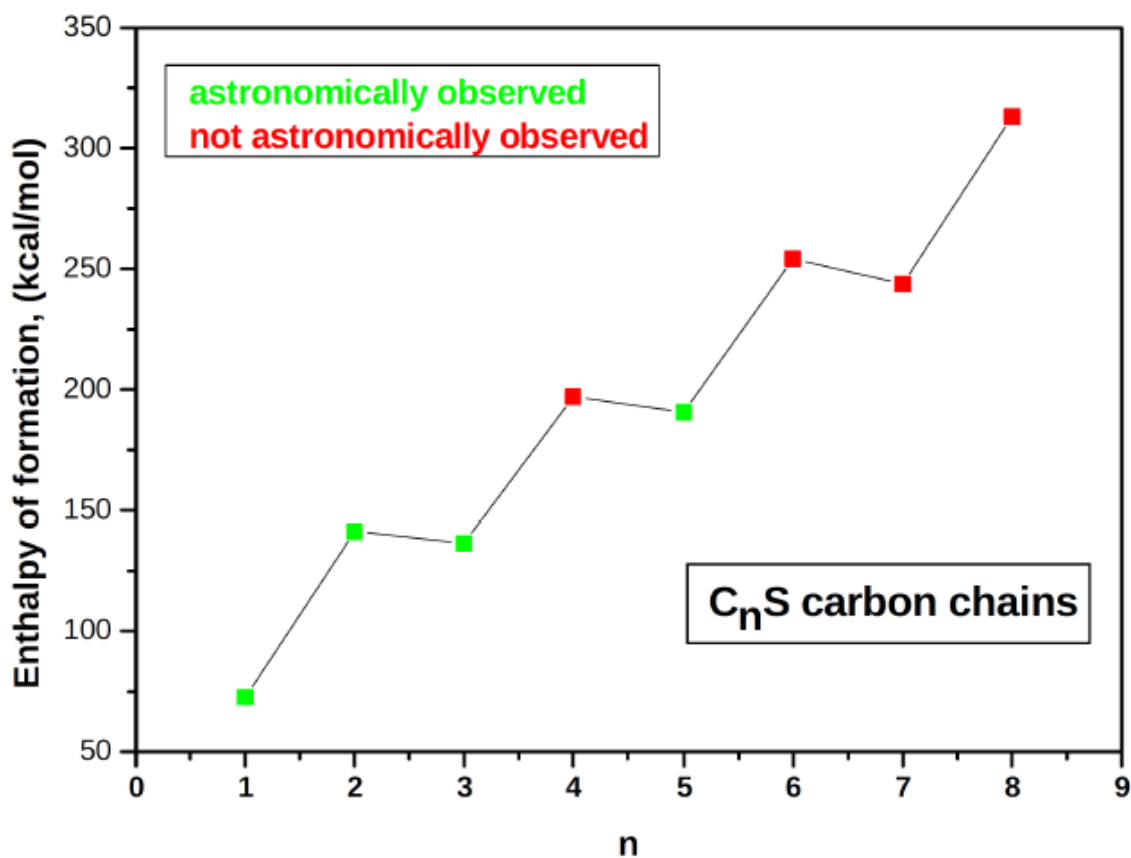

**Figure 3:** Plot showing the $\Delta_fH^O$ for $C_nS$ chain molecules

**Table 3:** $\Delta_fH^O$ for $C_nS$ chains and current astronomical status

| n | $\Delta_fH^O$ (kcal/mol) | Astronomical status |
|---|---|---|
| 1 | 72.404 | [a]Observed |
| 2 | 141.307 | [b]Observed |
| 3 | 136.292 | [c]Observed |
| 4 | 196.958 | Not observed |
| 5 | 190.684 | [d]Observed |
| 6 | 254.207 | Not observed |
| 7 | 243.688 | Not observed |
| 8 | 313.195 | Not observed |

[a]Penzias et al. 1971; [b]Saito et al. 1987; [c]Yamamoto et al.1987; [d]Cernicharo et al. 1987; [e]Agúndez et al. 2014

**3.4 $C_nN$ Chains:** The long carbon chain radicals like $C_nN$ have long been thought to be present in the interstellar and circumstellar media in detectable abundance and supposed to play an important role in astrophysics. A number of fundamental interstellar molecules were first detected in space before being observed in the gas phase in the terrestrial laboratory, these include $HCO^+$, $C_2H$, $HNC$, $HN_2^+$ (Guélin & Thaddeus, 1977) among others, thus most of them were termed "non-terrestrial". Cyanoethynyl radical, $C_3N$ is one of such molecules whose identification was guided by the ESR spectrum of a related species, $C_4H$ (butadinyl radical). The astronomical observations of $C_2N$ and $C_5N$ were hindered by lack of accurate spectroscopic data. Their astronomical observations became possible immediately after their laboratory microwave spectra were measured (Jefferts et al. 1970; Friberg et al. 1980; Guélin et al. 1998; Anderson & Ziurys 2014). Table 4 and Figure 4 indicate a constant gradual increase in the enthalpy of formation of the $C_nN$ chain molecules. A good correlation between the observations of the lower chains (except for the $C_4N$) with the enthalpy of formation is seen.

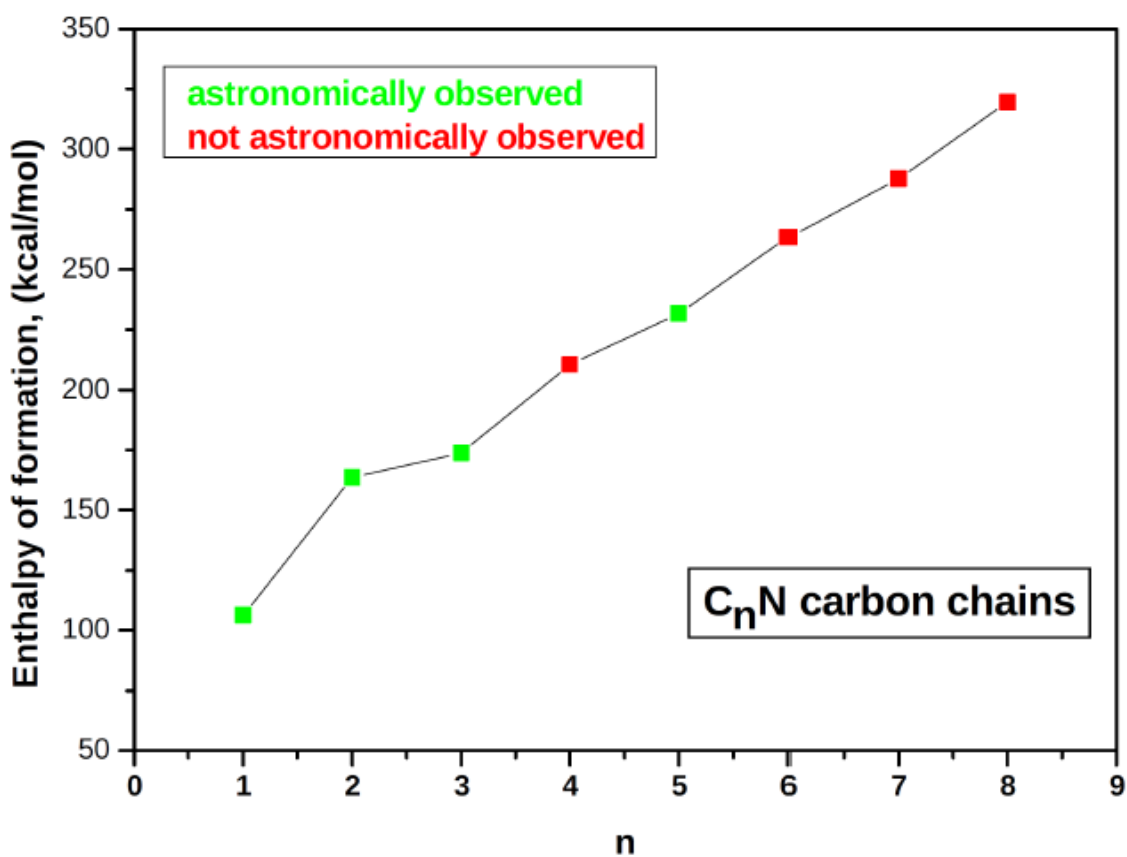

**Figure 4:** Plot showing the $\Delta_fH^O$ for $C_nN$ chain molecules

**Table 4:** $\Delta_fH^O$ for $C_nN$ chains and current astronomical status

| n | $\Delta_fH^O$ (kcal/mol) | Astronomical status |
|---|---|---|
| 1 | 106.413 | [a]Observed |
| 2 | 163.544 | [b]Observed |
| 3 | 173.701 | [c]Observed |
| 4 | 210.486 | Not observed |
| 5 | 231.723 | [d]Observed |
| 6 | 263.488 | Not observed |
| 7 | 287.663 | Not observed |
| 8 | 319.404 | Not observed |

[a]Jefferts et al. 1970; [b]Anderson & Ziurys 2014; [c]Friberg et al. 1980; [d]Guélin et al. 1998

**3. 5 $C_nSi$ Chains:** Every unique astronomical observation of molecule(s) in the interstellar or circumstellar medium has always been a concerted effort between the laboratory spectroscopists and the astrophysicists. The successful detection (after a number of failed attempts) of SiC in IRC+10216 (Cernicharo et al. 1989), came immediately after the successful measurements of its laboratory spectrum. Whether cyclic molecules are chemically less abundant than their related linear species in the interstellar or circumstellar medium has been the question for decades now, due to the fact that of about 200 known interstellar and circumstellar molecules, only about 10 are cyclic. However, of the 4 known $C_nSi$ molecules, 2 are cyclic. The ground state geometry of $SiC_2$ was shown to have a compact symmetric ($C_{2v}$) ring against the earlier believe that $SiC_2$ is linear. The proper elucidation of its ground-state geometry became instrumental for its astronomical observation (Thaddeus et al. 1984). Figure 5 shows the most stable geometry for the $SiC_3$ molecule which is the rhomboidal isomer. The experimentally determined bond lengths of c-$SiC_3$ by method of isotopic substitution measurement (Apponi et al. 1999) and the theoretically calculated bond lengths at the G4 level are in good agreements as shown in Figure 5A and 5B respectively. Table 5 and Figure 6 show the enthalpies of formation for all the $C_nSi$ molecules considered in this study, there is a steady increase in the enthalpy of formation value as the carbon chain increases except for the c-$C_2Si$. The observed molecules are also the most stable molecules (Cernicharo et al. 1989; Thaddeus et al.1984; Apponi et al. 1999; Ohishi et al. 1989). The enthalpy of formation for the linear $C_3Si$ molecule; 221.687 kcal/mol is about 10kcal/mol higher than the most stable isomer, which is the cyclic structure. The detection of more silicon-carbon molecules in IRC+10216 in particular (where all $C_nSi$ chains have been observed) and other astronomical sources await laboratory spectra of such molecules.

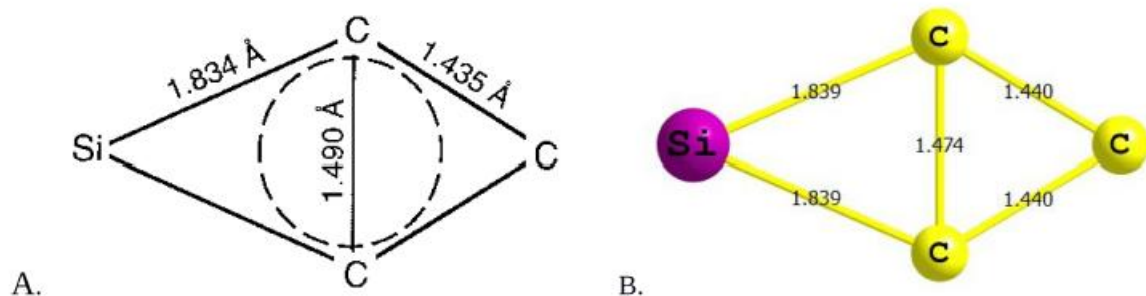

**Figure 5:** Experimental (A) and theoretical geometry (B) of c-SiC$_3$ molecule

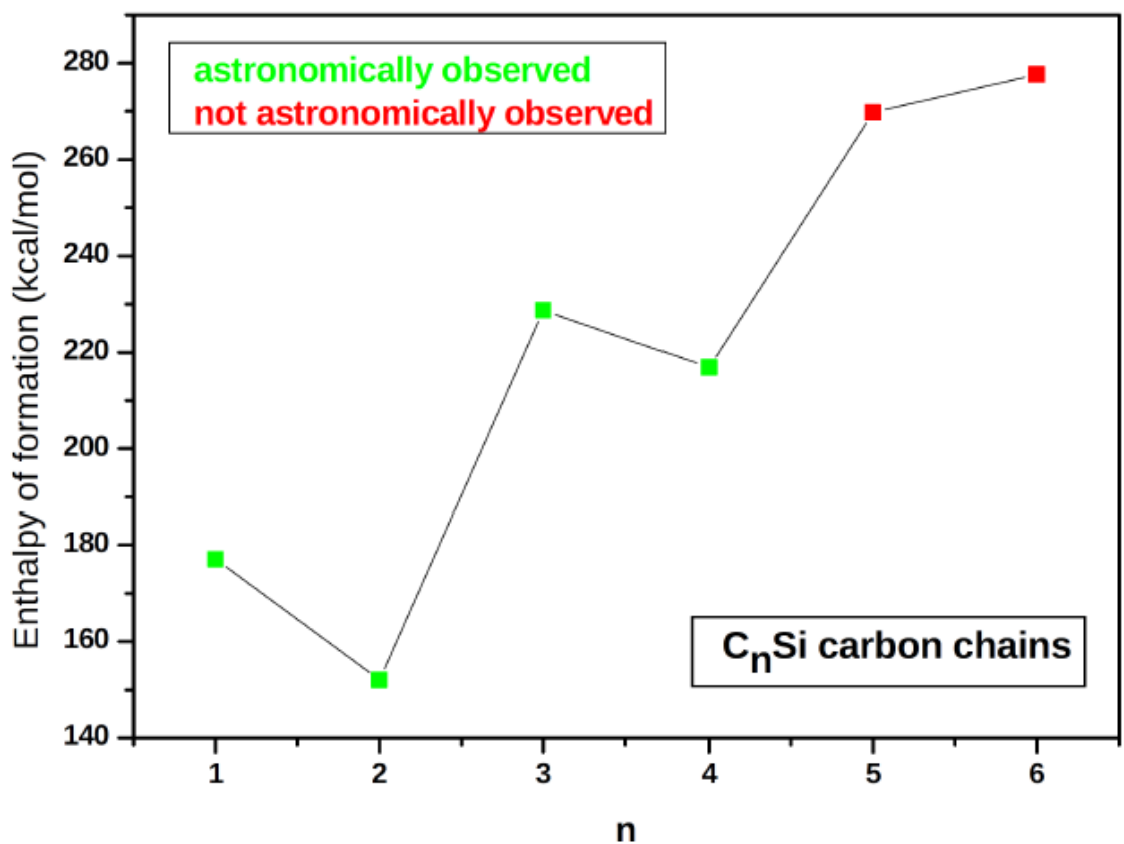

**Figure 6:** Plot showing the $\Delta_fH^O$ for C$_n$Si chain molecules

Table 5: $\Delta_fH^O$ for C$_n$Si chains and current astronomical status

| n | $\Delta_fH^O$ (kcal/mol) | Astronomical status |
|---|---|---|
| 1 | 177.113 | Observed (IRC+10216) |
| c-C$_2$Si | 151.989 | Observed (IRC+10216) |
| c-C$_3$Si | 228.773 | Observe (IRC+10216) |
| 4 | 216.930 | Observed (IRC+10216) |
| 5 | 269.727 | Not observed |
| 6 | 277.600 | Not observed |

[a]Cernicharo et al. 1989; [b]Thaddeus et al.1984; [c]Apponi et al. 1999; [d]Ohishi et al. 1989.

**3.6 HC$_n$N Chains:** The largest claimed interstellar linear molecule; the cyanopolyne, HC$_{11}$N belongs to the HC$_n$N chains. The enthalpies of formation of these molecules follow a unique trend with the odd number carbon chains being more stable than their preceding and the next even number carbon chains as shown in Table 6. For instance, HC$_3$N is more stable than both $HC_2N$ and $HC_4N$; $HC_5N$ is more stable than both $HC_4N$ and $HC_6N$, etc. Figure 7 depicts the zigzag pattern of the enthalpies of formation of the $HC_nN$ chains. Of the 8 molecules belonging to the $HC_nN$ carbon chain that have been detected in interstellar and circumstellar media (Snyder & Buhl 1971; Turner, 1971; Avery et al. 1976; Kroto et al., 1978; Broten et al., 1978; Guelin & Cernicharo, 1991; Bell et al., 1997; Cernicharo et al., 2004), only two ($HC_2N$ and $HC_4N$) are even number carbon chains while the remaining 6 are odd number carbon chains (HCN, $HC_3N$, $HC_5N$, $HC_7N$, $HC_9N$, $HC_{11}N$). A very good correlation between the observations of more odd number $HC_nN$ chains and their enthalpy of formation values in comparison to the even number $HC_nN$ chains is seen. The odd number $HC_nN$ are found to be more stable, and thus may be more easily observed as compared to the even number $HC_nN$ chains. From this analogy, $HC_{13}N$ is a more probable candidate for astronomical observation than both $HC_{12}N$ and $HC_{14}N$. Accurate laboratory rest frequencies of $HC_{13}N$ will be crucial for its astronomical searches and possible detection.

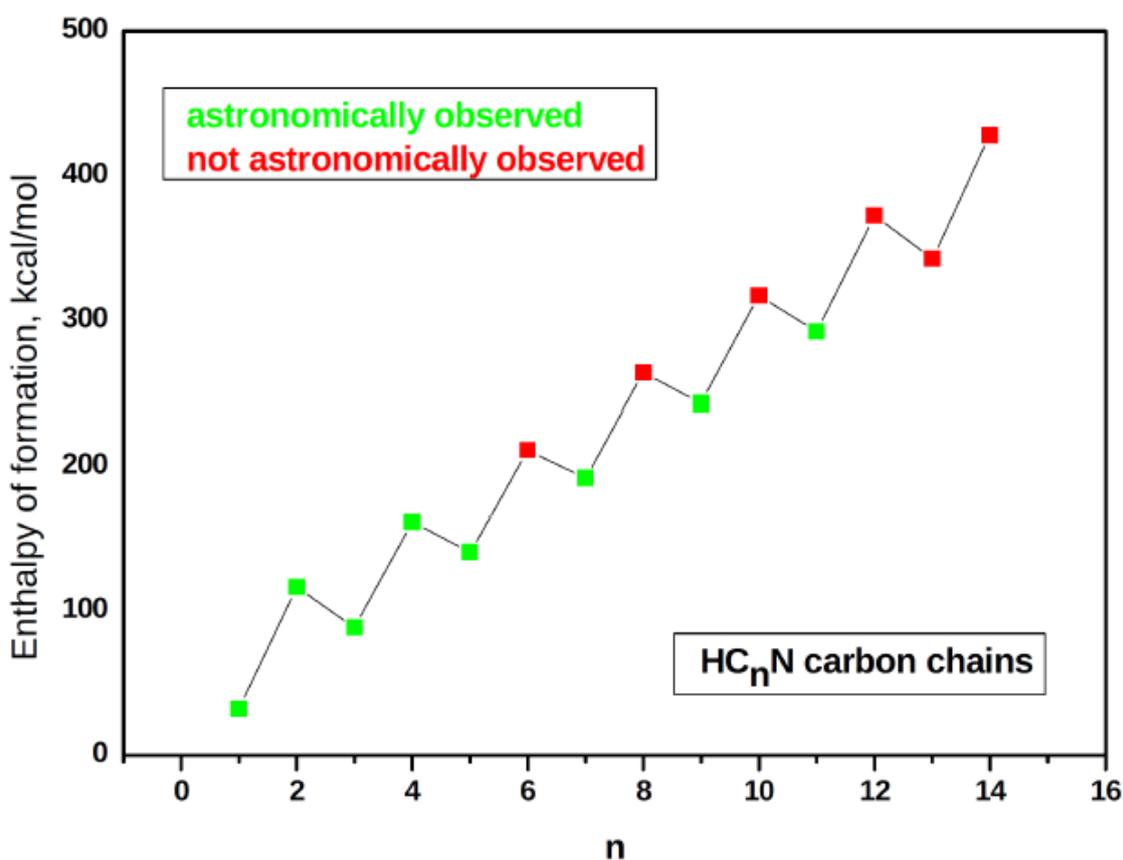

**Figure 7:** Plot showing the $\Delta_f H^O$ for $HC_nN$ chain molecules

**Table 6:** $\Delta_f H^O$ for $HC_nN$ chains and current astronomical status

| n | $\Delta_f H^O$ (kcal/mol) | Astronomical status |
|---|---|---|
| 1 | 32.192 | [a]Observed |
| 2 | 116.586 | [b]Observed |
| 3 | 88.738 | [c]Observed |
| 4 | 161.602 | [d]Observed |
| 5 | 140.566 | [e]Observed |
| 6 | 211.115 | Not observed |
| 7 | 191.824 | [f]Observed |
| 8 | 263.903 | Not oberved |
| 9 | 242.928 | [g]Observed |
| 10 | 317.217 | Not observed |
| 11 | 292.191 | [h]Observed |
| 12 | 372.551 | Not observed |
| 13 | 342.655 | Not observed |
| 14 | 428.028 | Not observed |

[a]Snyder & Buhl 1971, [b]Guelin & Cernicharo, 1991, [c]Turner, 1971, [d]Cernicharo et al., 2004, [e]Avery et al., 1976, [f]Kroto et al., 1978, [g]Broten et al., 1978, [h]Bell et al., 1997.

**3.7 $C_nH$ Chains:** Discoveries of cyanopolynes set the pace for the recognition of long chain molecules in space, such as $C_nH$, $C_nH^-$ etc. The acetylenic chain ($C_nH$) radical from n=1 to n=8 have all been observed in space, (Swings & Rosenfeld, 1937; Tucker et al., 1974; Thaddeus et al., 1985a; Guelin et al., 1978; Cernicharo et al., 1986; Suzuki et al., 1986; Cernicharo & Guelin, 1996; Guelin et al., 1997). The steady increase in the enthalpy of formation of these chains as the carbon chain increases is shown in Table 7 and Figure 8. Most of the $C_nH$ radicals could be termed as "non-terrestrial" as they were first observed in space using results from quantum chemical calculations prior to being studied in the laboratory. They are believed to grow steadily in the recombination phase of interstellar medium (Suzuki 1983). In the carbon rich astronomical sources such as TMC and IRC+10216 where all the known $C_nH$ chains radicals have been detected, successive members of the $C_nH$ chains (n>8) are more likely to exist.

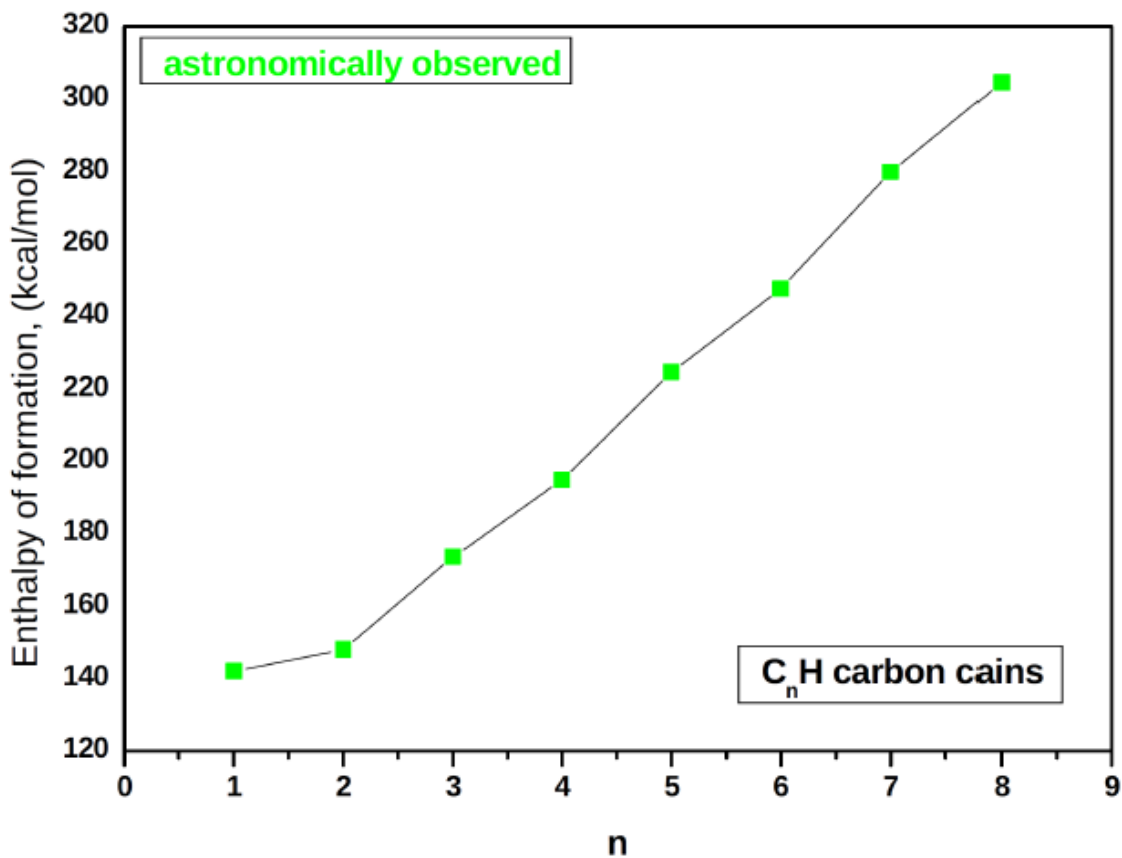

**Figure 8:** Plot showing the $\Delta_fH^O$ for $C_nH$ chain molecules

**Table 7:** $\Delta_fH^O$ for $C_nH$ chains and current astronomical status

| n | $\Delta_fH^O$ (kcal/mol) | Astronomical status | |
|---|---|---|---|
| 1 | 142.025 | [a]Observed | 142.025 |
| 2 | 148.023 | [b]Observed | 134.951 |
| 3 | 173.615 | [c]Observed | 173.615 |
| 4 | 194.790 | [d]Observed | 194.790 |
| 5 | 224.678 | [e]Observed | 224.678 |
| 6 | 247.560 | [f]Observed | 247.560 |
| 7 | 279.702 | [g]Observed | 279.702 |
| 8 | 304.480 | [h]Observed | 304.481 |

[a]Swings & Rosenfeld, 1937; [b]Tucker et al., 1974; [c]Thaddeus et al., 1985a; [d]Guelin et al., 1978; [e]Cernicharo et al., 1986; [f]Suzuki et al., 1986; [g]Guelin et al., 1997; [h]Cernicharo & Guelin, 1996

**3.8 $C_nP$ Chains:** Just like sulphur and oxygen, phosphorus (the 18th most cosmically abundant element) and nitrogen are in the same group in the periodic table. Whereas the similarity between the interstellar of sulphur and oxygen is well established with almost all S-containing interstellar molecules having the O-analogue as a known interstellar molecule, this is not the case in terms of the interstellar chemistry of N-containing molecules and the P-analogue with only very few known P-containing molecules as compared with the large number of N-containing interstellar/circumstellar molecules. This is due to the cosmic ratio of phosphorus and nitrogen as compared with the cosmic ratio of sulphur and oxygen. Table 8 and Figure 9 display the enthalpy of formation for all the $C_nP$ chains considered in this study. The only two astronomically observed molecules in this group; CP and $C_2P$ (Guelin et al., 1990, Halfen et al., 2008) are also the most stable, thus they may be detectable. Likewise the $C_nH$ chains, there is a corresponding increase in the enthalpy of formation value for every successive increase in the carbon chain.

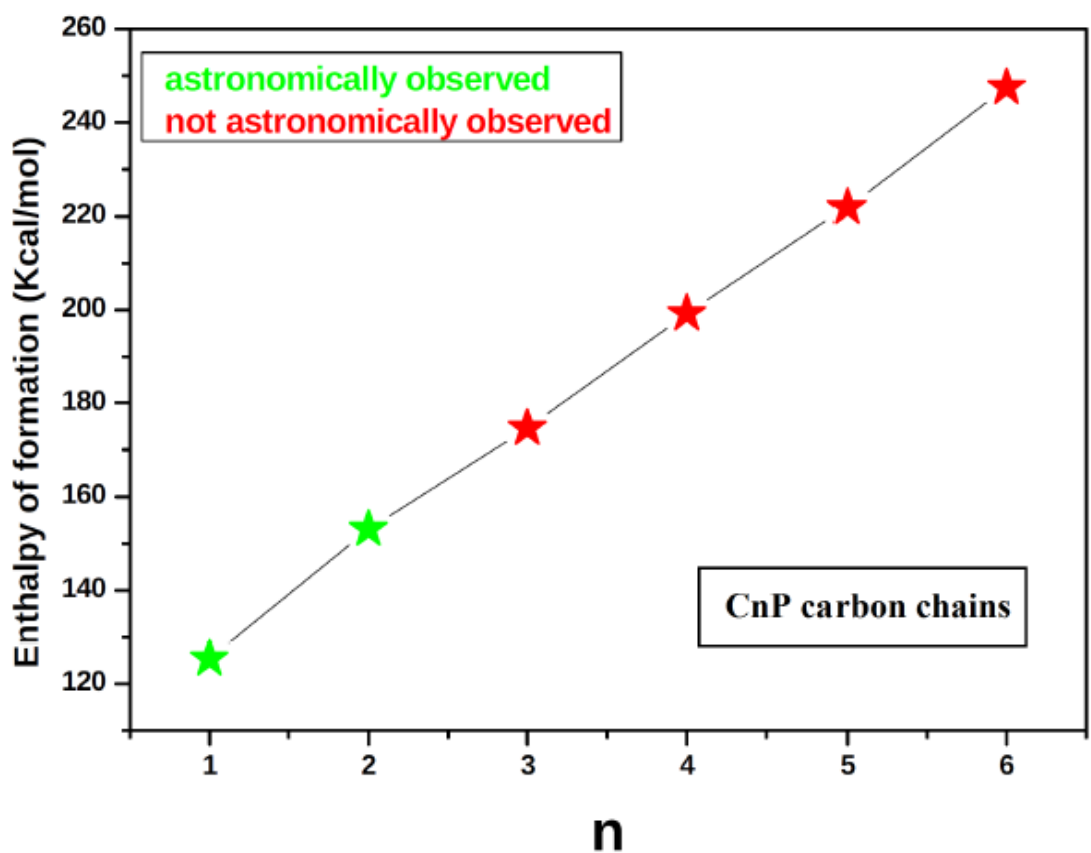

**Figure 9:** Plot showing the $\Delta_fH^O$ for $C_nP$ chain molecules

**Table 8:** $\Delta_fH^O$ for $C_nP$ chains and current astronomical status

| n | $\Delta_fH^O$ (kcal/mol) | Astronomical status |
|---|---|---|
| 1 | 125.335 | [a]Observed |
| 2 | 153.136 | [b]Observed |
| 3 | 174.674 | Not observed |
| 4 | 199.132 | Not observed |
| 5 | 221.889 | Not observed |
| 6 | 247.577 | Not observed |

[a]Guelin et al., 1990, [b]Halfen et al., 2008

**3.9 $H_2C_n$ Chains:** Interstellar acetylene is a well known starting material in the formation processes of many interstellar and circumstellar species (Rigway at al., 1976, Lacy et al., 1989). For the molecules with the empirical formula, $H_2C_3$, both the linear and the cyclic isomers have been observed in the Taurus molecular cloud, i.e., TMC-1 (Cernicharo et al., 1991, Thaddeus et al., 1985b). The cyclic isomer is now recognized as a common feature of many molecular clouds/astronomical sources. The polyacetylenic chains $C_4H_2$ and $C_6H_2$ have been observed in the proto–planetary nebula CRL 618 (Cernicharo et al., 2001) via their infrared vibrational spectra. The first three carbon chains in the $H_2C_n$ group have been astronomically observed as shown in Table 9 and Figure 10, the next most likely observable member of this group is not $H_2C_5$, neither is it $H_2C_7$, rather it is the $H_2C_6$. The even number carbon chains in $H_2C_n$ are more stable than their corresponding odd number carbon counterparts, thus probably accounting for the astronomical observation of more even carbon chains than the odd number carbon chains in this group.

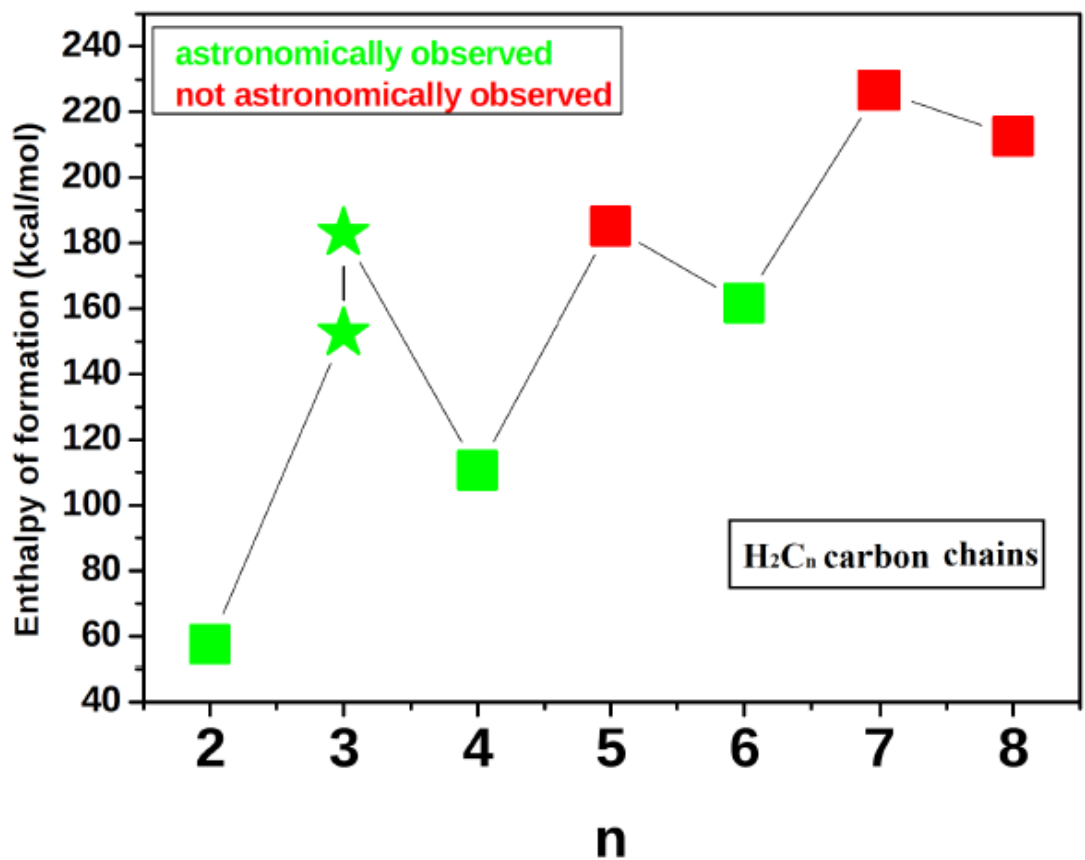

**Figure 10:** Plot showing the $\Delta_fH^O$ for $H_2C_n$ chain molecules

**Table 9:** $\Delta_fH^O$ for $H_2C_n$ chains and current astronomical status

| n | $\Delta_fH^O$ (kcal/mol) | Astronomical status |
|---|---|---|
| 2 | 57.612 | [a]Observed |
| 3(linear) | 152.590 | [b]Observed |
| 3 (cyclic) | 183.036 | [c]Observed |
| 4 | 110.685 | [d]Observed |
| 5 | 185.283 | Not observed |
| 6 | 161.678 | [d]Observed |
| 7 | 226.691 | Not observed |
| 8 | 212.684 | Not observed |

<sup>a</sup>Rigway at al., 1976, <sup>a</sup>Lacy et al., 1989; <sup>b</sup>Cernicharo et al., 1991, <sup>c</sup>Thaddeus et al., 1985b; <sup>d</sup>Cernicharo et al., 2001

**3.10 $C_nN^-$ Chains:** Among the ion molecules in interstellar and circumstellar media, the positive ions are known for long and observed while the negative ions are only known for less than a decade. Of the eight $C_nN^-$ chains considered here, only three have been astronomically observed; $CN^-$, $C_3N^-$ and $C_5N^-$ (Thaddeus et al., 2008, Cernicharo et al., 2008; Agúndez et al. 2010). The observed molecules (odd number carbon chains) are found to be more stable than the others. $CN^-$ is more stable than $C_2N^-$; $C_3N^-$ is more stable than both $C_2N^-$ and $C_4N^-$; $C_5N^-$ is more stable than both $C_4N^-$ and $C_6N^-$ (Table 10). Figure 11 depicts the plot of the enthalpy of formation of these molecules which shows a zigzag pattern as noted in the case of the $HC_nN$ chains. Following the order of their astronomical observations, $C_7N$ can be considered as the next probable candidate for astronomical observation as it is more stable than both the **$C_6N^-$** and the $C_8N$-.

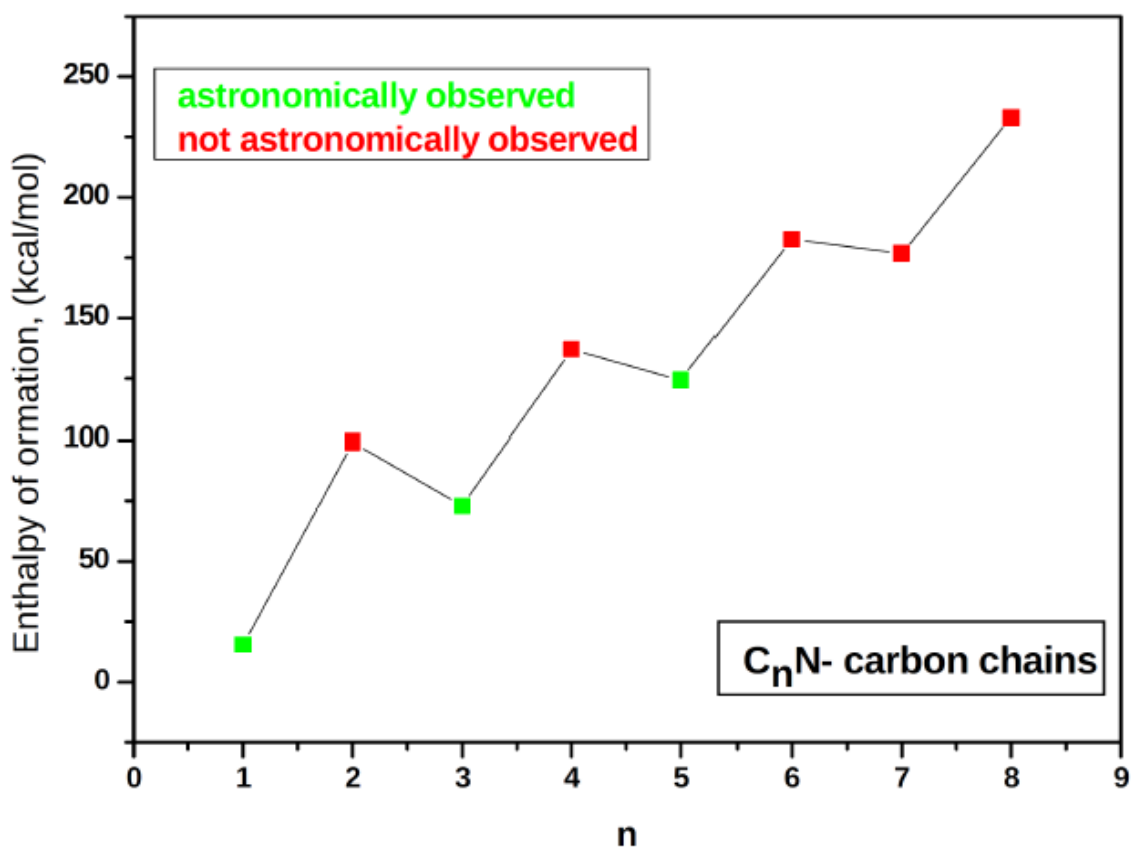

**Fig. 11.** Plot showing the $\Delta_fH^O$ for $C_nN$- chain molecules

**Table 10:** $\Delta_fH^O$ for $C_nN^-$ chains and current astronomical status

| n | $\Delta_fH^O$ (kcal/mol) | Astronomical status |
|---|---|---|
| 1 | 15.459 | <sup>a</sup>observed |
| 2 | 99.183 | Not observed |
| 3 | 72.910 | [b]Observed |
| 4 | 137.056 | Not observed |
| 5 | 124.358 | [c]Observed |
| 6 | 182.629 | Not observed |
| 7 | 176.752 | Not observed |
| 8 | 232.983 | Not observed |

[a]Agúndez et al. 2010; [b]Thaddeus et al., 2008, [c]Cernicharo et al., 2008

**3.11 $C_nH^-$ Chains:** Molecular anions have long been predicted to be detectable in astronomical sources, and are currently estimated by chemical formation models to be in high abundance than can be compared to their neutral counterparts. The first known interstellar molecular anion, $C_6H^-$ was shown to be the carrier of the unidentified series of lines IRC+10216 via laboratory measurements of its rotational spectrum. The astronomical observations of $C_4H^-$ and $C_8H^-$ came immediately after the successful measurements of their rotational spectra (Cernicharo et al., 2007; McCarthy et al., 2006; Brunken et al., 2007a). Theoretically estimated enthalpy of formation for the $C_nH^-$ (n=1 to 10) chains using the G4 composite method is presented in Table 11, the same is also displayed in Figure 12. The even number carbons chains are more stable than the preceding and the next odd number carbon chains; $C_2H^-$ is more stable than $CH^-$ and $C3H$, $C_4H^-$ is more stable than $C_3H^-$ and $C_5H^-$, and so on. The rotational spectrum of $C_2H^-$ is known (Brünken et al., 2007b), its astronomical observation is a function of time. The three astronomically detected $C_nH^-$ are all even number carbon chains; these are the most stable which might accounts for their observation as compared to the preceding and the next odd number chains in each case which have not been observed.

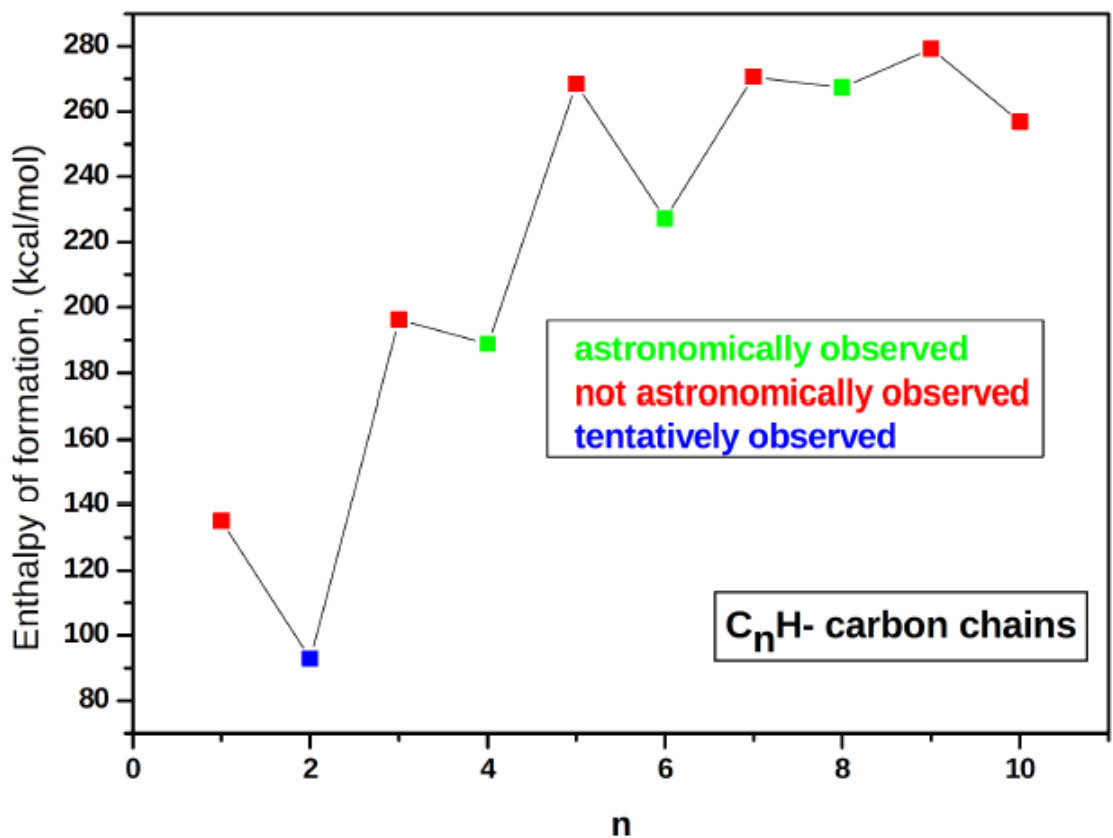

**Figure 12:** Plot showing the $\Delta_f H^O$ for $C_nH^-$ chain molecules

**Table 11:** $\Delta_f H^O$ for $C_nH^-$ chains and current astronomical status

| n | $\Delta_f H^O$ (kcal/mol) | Astronomical status |
|---|---|---|
| 1 | 135.131 | Not observed |
| 2 | 92.858 | [d]Tentatively observed |
| 3 | 196.282 | Not observed |
| 4 | 188.822 | [a]Observed |
| 5 | 268.466 | Not observed |
| 6 | 227.126 | [b]Observed |
| 7 | 270.745 | Not observed |
| 8 | 267.406 | [c]Observed |
| 9 | 279.286 | not observed |
| 10 | 256.98 | not observed |

[a]Cernicharo et al., 2007; [b]McCarthy et al., 2006; [c]Brunken et al., 2007; [d]Ag´undez *et al.* 2010 and references therein.

Table 12 gives the summary of all the different carbon chains that have been considered in this study, it is obvious from the table that in each group, the number of the odd and even number carbon chains observed depend on the trend of the enthalpy of formation of the molecules in that group. The odd number carbon chains are found to be more stable than their corresponding even number carbon chains in most of the group (Figure 13); this might has resulted in the observation of more interstellar odd number (≈60%) carbon chains than the even number carbon chains.

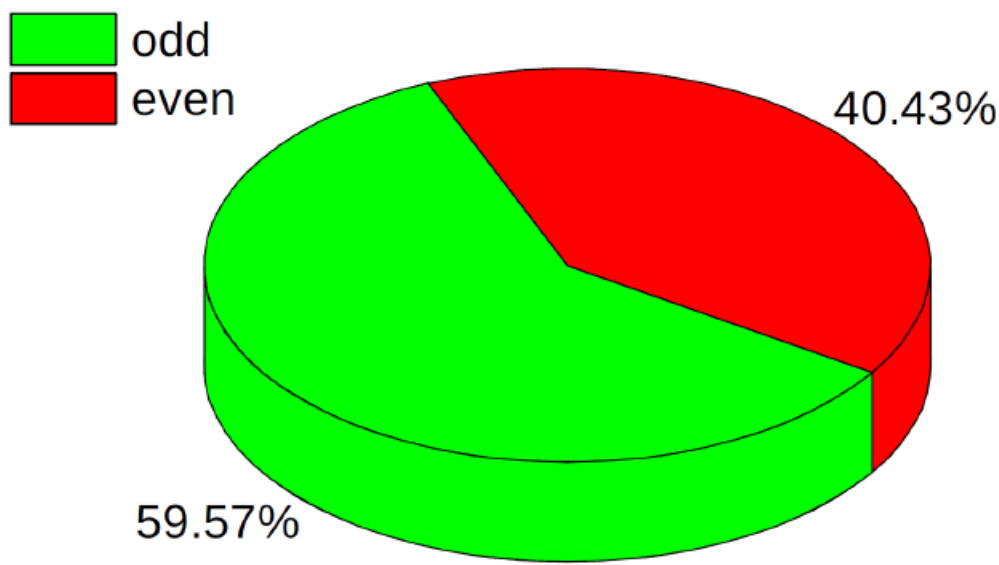

**Figure13:** Percentage of detected odd and even number carbon chains

**Table 12:** Carbon chains that have been considered in this study

| Chain | Total number considered | Odd observed | Even observed |
|---|---|---|---|
| $C_n$ | 7 | 2 | 1 |
| $C_nO$ | 6 | 2 | 1 |
| $C_nS$ | 8 | 3 | 1 |
| $C_nN$ | 8 | 3 | 1 |
| $C_nSi$ | 6 | 2 | 2 |
| $HC_nN$ | 14 | 6 | 2 |
| $C_nH$ | 8 | 4 | 4 |
| $C_nP$ | 6 | 1 | 1 |
| $C_nN^-$ | 8 | 3 | 0 |
| $C_nH^-$ | 10 | 0 | 3 |
| $H_2C_nN$ | 8 | 2 | 3 |
| Total | 89 | 28 | 19 |

## 3.12 Effect of Kinetics on the Formation of Carbon Chains

The effect of kinetics in the formation of carbon chain species both in the terrestrial laboratory and in the interstellar medium cannot be overemphasized. Though there is hardly a consensus as to how any molecule (especially the complex ones with six atoms and above) is formed in the interstellar medium, but the $C_2$ addition is the widely claimed route for the formation of the linear carbon chains. In the terrestrial laboratory, unsaturated molecules like acetylene, diacetylene, etc., are used as the source of the **$C_2$** in producing the carbon chains while in the interstellar medium, acetylene which produces the $C_2$ is generally accepted as the starting material for the formation of these carbon chains (Thaddeus et al., 1998; McCarthy et al., 1997; Cherchneff et al., 1993; Cherchneff and Glassgold, 1993). With respect to the carbon chains discussed here, a good correlation between the enthalpy of formation and observed abundances has been seen. For instance, in the $C_nO$ chains where the odd numbered carbon chains are more stable than their progressive even numbered chains, $C_2$ addition to the lower chains to form the higher chains clearly favours the odd numbered chains as compared to their corresponding even numbered chains. A $C_2$ addition to CO to form $C_3O$ is likely to have a lower barrier as compared to $C_2$ addition to $C_2O$ to form $C_4O$, which could account for the delayed observation of $C_4O$. This is also applicable to the $C_nS$ chains, a $C_2$ addition to $C_3S$ to form $C_5S$ (astronomically observed) is more favourable than a $C_2$ addition to $C_2S$ to form $C_4S$ (yet to be observed). And this goes on to the higher members of the chains. Also, in the $HC_nN$ and $C_nN^-$ chains, $C_2$ addition to form higher members of the chains is likely to favour the odd numbered carbon chains as compared to the even numbered chains which explains the trends in the astronomical observation of these species. In the case of $C_nSi$, $H_2C_n$, and $C_nH^-$ chains where the even numbered carbon chains are more stable than their corresponding odd numbered chains, it is crystal clear that $C_2$ addition to the lower members of the chains is more likely to favours the even numbered carbon chains as compared to their odd numbered counterparts. For instance, a $C_2$ addition to $HC_5N$ to form $HC_7N$ is could be more favourable than $C_2$ addition to $HC_4N$ to form $HC_6N$ which has not been detected. Correlation between the effects of kinetics in the formation of these species is seen even for the chains with a linear increase in the enthalpy of formation like the $C_nH$, $C_nP$, $C_nN$ and $C_n$ chains.

**3.13 Other Studies Supporting the Above Results:**

The correlation between the thermodynamics and observation of the linear interstellar carbon chains had been discussed in the previous sections. Here, it is also correlated with the chemical reactivity indices such as, global hardness ($\eta$) and softness (S). Soft species are more reactive and unstable in comparison to the hard species. Larger values of S indicate that the species are more unstable and larger values of $\eta$ indicates that the species are more stable. All the parameters are pointed out in Table 13. For the simplicity in Fig. 14, we have shown the softness values for the carbon chain species.

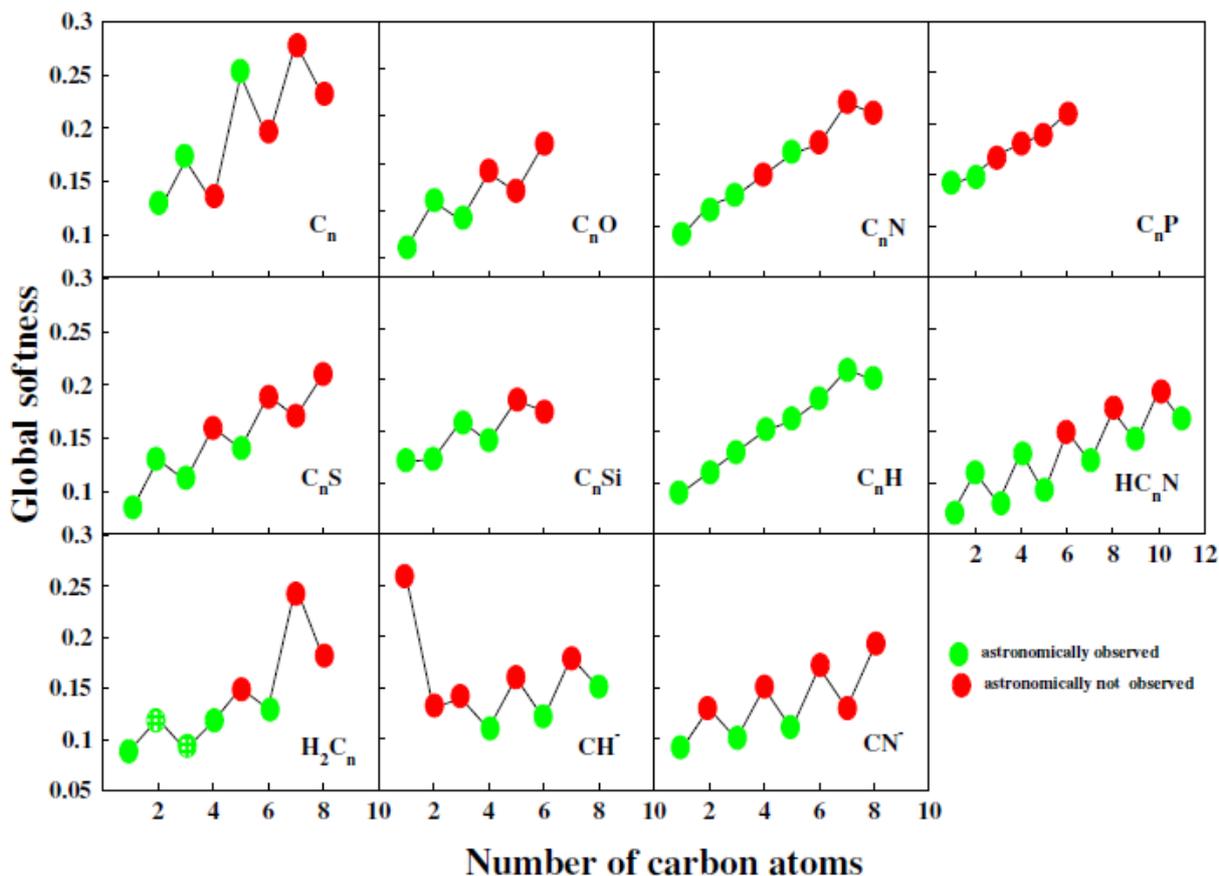

**Figure 14:** Variation of global softness with the number of carbon atoms present in carbon chains.

3.13.1 Chemical modelling:

Need for chemical modelling is ever increasing because of the recent development of various sensitive ground based and space-borne facilities as well as the vast data produced by the low temperature laboratory experiments. Chemical modelling plays a very important role to understand the empirical outcome of the observational results. We use our large gas-grain chemical network (Das et al. 2013, 2015a, 2015b, Majumdar et al. 2014a, 2014b) for the purpose of chemical modelling. We assume that gas and grains are coupled through accretion and thermal/non-thermal/cosmic ray evaporation processes. Details of these processes and chemical network were already presented in Das et al. (2015a, 2015b). Gas phase chemical network principally adopted from UMIST 2006 (Woodall et al. 2007) database. Since UMIST 2006 database does not consider some of the carbon chain molecules, we include their reactions from UMIST 2012 network (McElroy et al. 2012). These databases contain different assorted reaction pathways and rate coefficients from the literature and therefore our calculated thermodynamical parameters do not seem to play any role inside this modelling.

**3.13.2 Physical aspects**: Chemical enrichment of a molecular cloud is interrelated with the surrounding physical conditions and age of the cloud. Depending on the densities ($n_H$), temperatures (T) and visual extinction ($A_V$) parameters, Snow & McCall (2006) classified various types of clouds. Das et al. (2016) also assumed these classifications in their model. For the diffuse cloud region, they considered that $n_H$ may vary between $100 - 500$ cm$^{-3}$, T may vary in between $30 - 100$ K and $A_V = 0.2$, for the translucent cloud region, we used $n_H = 501 - 10000$ cm$^{-3}$, T = $15 - 30$ K, $A_V = 1 - 2$ and for the dense cloud region, we used $n_H = 10001 - 1000000$ cm$^{-3}$, T = 10 K and $A_V$ = 5-10. In order to consider more realistic condition, they considered different temperatures and extinction parameters for various regions of clouds. They used constant slopes for T and $A_V$ in respective number density windows. In diffuse and translucent clouds regions, gas and grains are not well coupled. So, temperature between these two phases might vary. However, in dense cloud regions gas and grains are strongly coupled and the temperature would be more or less the same for both the phases. They assumed that the grain temperature is always fixed at 10 K for all the clouds and for the gas temperature ($T_{gas}$), we assume $T_{gas}$= T. Here we present the results for the dense and translucent cloud regions. Following Das et al. (2015a), we use $n_H$=2 x $10^4$ cm$^{-3}$, $T_{gas}$=10K, $T_{gr}$=10K, $A_V$=5.05 for dense cloud and $n_H$=5 x $10^3$ cm$^{-3}$, $T_{gas}$=22.89 K, $T_{gr}$=10K, $A_V$=1.47 for translucent cloud.

We assume that initially most of the hydrogens were in the form of molecular hydrogen. Our adopted initial condition is shown in Table 14. The choice of our initial condition was based on the study of Leung, Herbst & Huebner (1984). These initial low metalicity elemental abundances are often used to calculate molecular abundances in cold, dense clouds. Gas phase species are allowed to accrete on the grains with a sticking coefficient of 0.3. Surface species are allowed to migrate through the grain surface depending upon their binding energies. Binding energies of the surface species are taken from some past studies (Allen & Robinson 1977; Tielens & Allamandola 1987; Hasegawa & Herbst 1993; Hasegawa, Herbst & Leung 1992).

**Table13:** Global reactivity indices for the carbon chain molecules

| Carbon Chains | Electron affinity (ev) | Ionization potential (ev) | Global hardness ($\eta$) | Global softness (S) | Electrophilicity index ($\omega$) |
|---|---|---|---|---|---|
| $C_2$ | 4.315589604 | 12.38118018 | 8.06559057 | 0.12398348 | 4.320547246 |
| $C_3$ | 4.299943058 | 10.03321861 | 5.73327555 | 0.17442036 | 4.479104517 |
| $C_4$ | 3.223841626 | 10.48334934 | 7.25950771 | 0.13775039 | 3.23518983 |
| $C_5$ | 5.061235976 | 8.983784335 | 3.92254836 | 0.25493631 | 6.286174748 |
| $C_6$ | 4.573961705 | 9.574978885 | 5.00101718 | 0.19995932 | 5.003795043 |
| $C_7$ | 5.054541975 | 8.581572853 | 3.52703088 | 0.28352459 | 6.589948956 |
| $C_8$ | 4.638425476 | 8.973879391 | 4.33545392 | 0.23065636 | 5.342429173 |
|  |  |  |  |  |  |
| CO | -1.15085110 | 14.19743128 | 15.3482824 | 0.06515387 | 1.386256536 |
| $C_2O$ | 2.34194785 | 11.3391474 | 8.99719955 | 0.1111457 | 2.600425363 |
| C$_3$O | 1.07569326 | 11.91692674 | 10.8412335 | 0.09224043 | 1.946367251 |
| C$_4$O | 2.9618232 | 10.00853788 | 7.04671468 | 0.1419101 | 2.984196789 |
| C$_5$O | 2.124637729 | 10.36919758 | 8.24455985 | 0.1212921 | 2.366650306 |
| C$_6$O | 3.324768653 | 9.175651811 | 5.85088316 | 0.17091437 | 3.338395837 |
| | | | | | |
| CH | 0.120138265 | 11.00240651 | 10.8822682 | 0.09189261 | 1.421015815 |
| C$_2$H | 3.111703509 | 11.56853937 | 8.45683586 | 0.11824754 | 3.185433869 |
| C$_3$H | 1.926647692 | 9.361641627 | 7.43499393 | 0.1344991 | 2.142326491 |
| C$_4$H | 3.47244484 | 10.36144233 | 6.88899749 | 0.145159 | 3.472501523 |
| C$_5$H | 2.593898064 | 8.655995991 | 6.06209793 | 0.16495939 | 2.609660009 |
| C$_6$H | 3.647169143 | 9.20098561 | 5.55381647 | 0.18005636 | 3.715352351 |
| C$_7$H | 2.975782641 | 7.758237977 | 4.78245534 | 0.20909761 | 3.01150744 |
| C$_8$H | 3.823907089 | 8.65754704 | 4.83363995 | 0.20688343 | 4.028710732 |
| | | | | | |
| CN | 4.962404226 | 15.61593357 | 10.6535293 | 0.09386561 | 4.968634956 |
| C$_2$N | 2.940870434 | 11.09087072 | 8.15000029 | 0.12269938 | 3.019781486 |
| C$_3$N | 4.493470428 | 12.46142655 | 7.96795612 | 0.1255027 | 4.509759578 |
| C$_4$N | 3.312577953 | 9.828969953 | 6.516392 | 0.15345915 | 3.312804873 |
| C$_5$N | 4.888198779 | 10.5489832 | 5.66078442 | 0.17665396 | 5.262225376 |
| C$_6$N | 3.605127553 | 9.117174544 | 5.51204699 | 0.1814208 | 3.67052773 |
| C$_7$N | 4.971356771 | 9.555305054 | 4.58394828 | 0.21815255 | 5.7544253 |
| C$_8$N | 3.803988355 | 8.626553273 | 4.82256492 | 0.20735854 | 4.005087743 |
| | | | | | |
| CP | 3.624420425 | 10.72087752 | 7.09645709 | 0.14091539 | 3.624829446 |
| C$_2$P | 2.731533249 | 9.229122182 | 6.49758893 | 0.15390324 | 2.752122361 |
| C$_3$P | 3.658761193 | 9.590434952 | 5.93167376 | 0.16858648 | 3.699234095 |
| C$_4$P | 3.154316538 | 8.617627938 | 5.4633114 | 0.18303917 | 3.170665797 |
| C$_5$P | 3.821539698 | 8.940409387 | 5.11886969 | 0.19535563 | 3.977131533 |
| C$_6$P | 3.431410071 | 8.207797269 | 4.7763872 | 0.20936326 | 3.545335153 |
| | | | | | |
| HCN | -0.70335987 | 13.50925535 | 14.2126152 | 0.07036003 | 1.442301049 |
| HC$_2$N | 1.977478559 | 10.69984312 | 8.72236456 | 0.11464781 | 2.303195479 |
| HC$_3$N | -0.05679016 | 11.341542 | 11.3983322 | 0.08773213 | 1.396537913 |
| HC$_4$N | 2.023765125 | 9.568992381 | 7.54522726 | 0.13253411 | 2.226441001 |
| HC$_5$N | 0.9548475 | 11.55373638 | 10.5988889 | 0.09434951 | 1.845295677 |
| HC$_6$N | 2.439038072 | 8.919565466 | 6.48052739 | 0.15430843 | 2.488568184 |
| HC$_7$N | 1.592246982 | 9.470487167 | 7.87824019 | 0.1269319 | 1.941805593 |
| HC$_8$N | 2.732295168 | 8.471257899 | 5.73896273 | 0.17424752 | 2.733934842 |
| HC$_9$N | 2.030105378 | 8.964491463 | 6.93438609 | 0.14420887 | 2.179016971 |
| HC$_{10}$N | 2.90070643 | 8.139006888 | 5.23830046 | 0.19090161 | 2.908273187 |
| HC$_{11}$N | 2.433160413 | 8.604403205 | 6.17124279 | 0.16204191 | 2.467651466 |
| | | | | | |
| H$_2$C$_2$ | 0.677808389 | 11.27155432 | 10.5937459 | 0.09439532 | 1.684806183 |
| l-H$_2$C$_3$ | 1.945042588 | 10.29523704 | 8.35019445 | 0.11975769 | 2.242828688 |
| c-H$_2$C$_3$ | -1.08125684 | 9.15234512 | 10.2336010 | 0.0977517 | 0.79575879 |
| H$_2$C$_4$ | 1.859653262 | 10.34897952 | 8.48932626 | 0.11779498 | 2.194678204 |
| H$_2$C$_5$ | 2.618905327 | 9.231217459 | 6.61231213 | 0.15123303 | 2.654620033 |
| H$_2$C$_6$ | 2.548155726 | 10.02097349 | 7.47281776 | 0.13381833 | 2.642627825 |
| H$_2$C$_7$ | 3.056083438 | 6.991176246 | 3.93509281 | 0.25412361 | 3.206640608 |
| H$_2$C$_8$ | 2.988626415 | 8.396290533 | 5.40766412 | 0.18492273 | 2.996125755 |
| CSi | 2.293974178 | 10.54732331 | 8.25334913 | 0.12116293 | 2.497454647 |
| $C_2Si$ | 1.678207747 | 9.765921174 | 8.08771343 | 0.12364434 | 2.024182854 |
| $C_3Si$ | 2.66475651 | 9.066534157 | 6.40177765 | 0.15620661 | 2.687206357 |
| $C_4Si$ | 2.343199574 | 9.39489394 | 7.05169437 | 0.14180989 | 2.442371155 |
| $C_5Si$ | 3.048681941 | 8.506959236 | 5.45827729 | 0.18320799 | 3.058035434 |
| $C_6Si$ | 2.784622661 | 8.774501573 | 5.98987891 | 0.16694828 | 2.788314986 |
| | | | | | |
| CS | 0.299352446 | 11.50772193 | 11.2083695 | 0.08921904 | 1.554719952 |
| $C_2S$ | 2.689736562 | 10.40672208 | 7.71698552 | 0.12958428 | 2.778242044 |
| $C_3S$ | 1.702779628 | 10.75483733 | 9.0520577 | 0.11047212 | 2.143051687 |
| $C_4S$ | 3.159758815 | 9.441370985 | 6.28161217 | 0.1591948 | 3.159787407 |
| $C_5S$ | 2.481569466 | 9.63361942 | 7.15204995 | 0.13982005 | 2.565309993 |
| $C_6S$ | 3.452934277 | 8.804488519 | 5.35155424 | 0.1868616 | 3.509363981 |
| $C_7S$ | 2.959483021 | 9.001607792 | 6.04212477 | 0.16550469 | 2.95979682 |
| $C_8S$ | 3.693210806 | 8.368290018 | 4.67507921 | 0.21390012 | 3.889768375 |
| | | | | | |
| $CN^-$ | -6.72932108 | 4.962404226 | 11.6917253 | 0.08553058 | 0.033378256 |
| $C_2N^-$ | -4.94071675 | 2.940870434 | 7.88158719 | 0.126878 | 0.06342925 |
| $C_3N^-$ | -5.60834808 | 4.493470428 | 10.1018185 | 0.09899208 | 0.015380302 |
| $C_4N^-$ | -3.36525919 | 3.312577953 | 6.67783715 | 0.14974908 | 5.19501E-05 |
| $C_5N^-$ | -3.91196313 | 4.888198779 | 8.80016191 | 0.11363427 | 0.013537195 |
| $C_6N^-$ | -2.21244886 | 3.605127553 | 5.81757642 | 0.17189289 | 0.04167444 |
| $C_7N^-$ | -2.62399385 | 4.971356771 | 7.59535063 | 0.13165949 | 0.090682329 |
| $C_8N^-$ | -1.34579346 | 3.803988355 | 5.14978182 | 0.19418298 | 0.146674226 |
| | | | | | |
| $CH^-$ | -3.69620405 | 0.120138265 | 3.81634232 | 0.262031 | 0.418864631 |
| $C_2H^-$ | -4.15879760 | 3.111703509 | 7.27050111 | 0.1375421 | 0.018850249 |
| $C_3H^-$ | -5.00950713 | 1.926647692 | 6.93615483 | 0.1441721 | 0.171276856 |
| $C_4H^-$ | -5.56902763 | 3.47244484 | 9.04147247 | 0.11060145 | 0.06077079 |
| $C_5H^-$ | -3.65484275 | 2.593898064 | 6.24874082 | 0.16003224 | 0.022516609 |
| $C_6H^-$ | -4.06075498 | 3.647169143 | 7.70792413 | 0.12973662 | 0.002773984 |
| $C_7H^-$ | -2.53308061 | 2.975782641 | 5.50886326 | 0.18152565 | 0.00444704 |
| $C_8H^-$ | -2.99548368 | 3.823907089 | 6.81939077 | 0.14664067 | 0.012579667 |

**Table 14:** Initial abundances

| Species | Abundance w.r.t. total H |
|---|---|
| H2 | $5.00 \times 10^{-01}$ |
| He | $1.00 \times 10^{-01}$ |
| N | $2.14 \times 10^{-05}$ |
| O | $1.76 \times 10^{-04}$ |
| H3+ | $1.00 \times 10^{-11}$ |
| C+ | $7.30 \times 10^{-05}$ |
| S+ | $8.00 \times 10^{-08}$ |
| Si+ | $8.00 \times 10^{-09}$ |
| Fe+ | $3.00 \times 10^{-09}$ |
| Na+ | $2.00 \times 10^{-09}$ |
| Mg+ | $7.00 \times 10^{-09}$ |
| P+ | $3.00 \times 10^{-09}$ |
| Cl+ | $4.00 \times 10^{-09}$ |
| e- | $7.31 \times 10^{-05}$ |
| HD | $1.60 \times 10^{-05}$ |

Chemical evolution of the carbon chain molecules are shown in Fig. 15 (for dense cloud region) and Fig. 16 (for translucent cloud region). For the dense cloud region, gas phase abundances of the carbon chain molecules are decreased at the latter stages (~$10^6$ years). This happened due to the heavy depletion of the gas phase species. Since, depletion time scale is inversely proportional to the number density of a cloud, for translucent cloud region, no such depletion is prominent around the latter stages. Abundance of interstellar species is not driven by thermodynamic stability of that species, it primarily depends on some other factors like: formation and destruction rate of the species, relative concentration of the interacting species, size of the species (here it depends on the number of carbons in it), etc.. For example, for the formation of $C_nX$, it would require some extra reaction steps in comparison to the formation of $C_{n-1}X$). So, while comparing the abundances of any species with the effect of thermodynamics, it should be with appropriate weighted average. Peak and final abundances of these species are shown in Table 15 after $10^6$ years. Keeping in mind all these aspects, abundances might be correlated with the energy and stability values of several carbon chain species. Few prominent examples of these correlations could be seen from Fig. 15, Fig. 16 and Table 15. Some examples are;

a) From Fig. 12, we can see that $C_6H^-$ is more stable than $C_5H^-$, $C_7H^-$, $C_8H^-$, $C_9H^-$, $C_{10}H^-$ and Fig. 15, Fig. 16 and Table 15, we find that $C_6H^-$ is comparatively more abundant as expected.

b) From Fig. 10, it is evident that $H_2C_6$ is more stable than $H_2C_5$, $H_2C_7$, $H_2C_8$. Same thing is also prominent from Fig.16, Fig. 16 and Table 15.

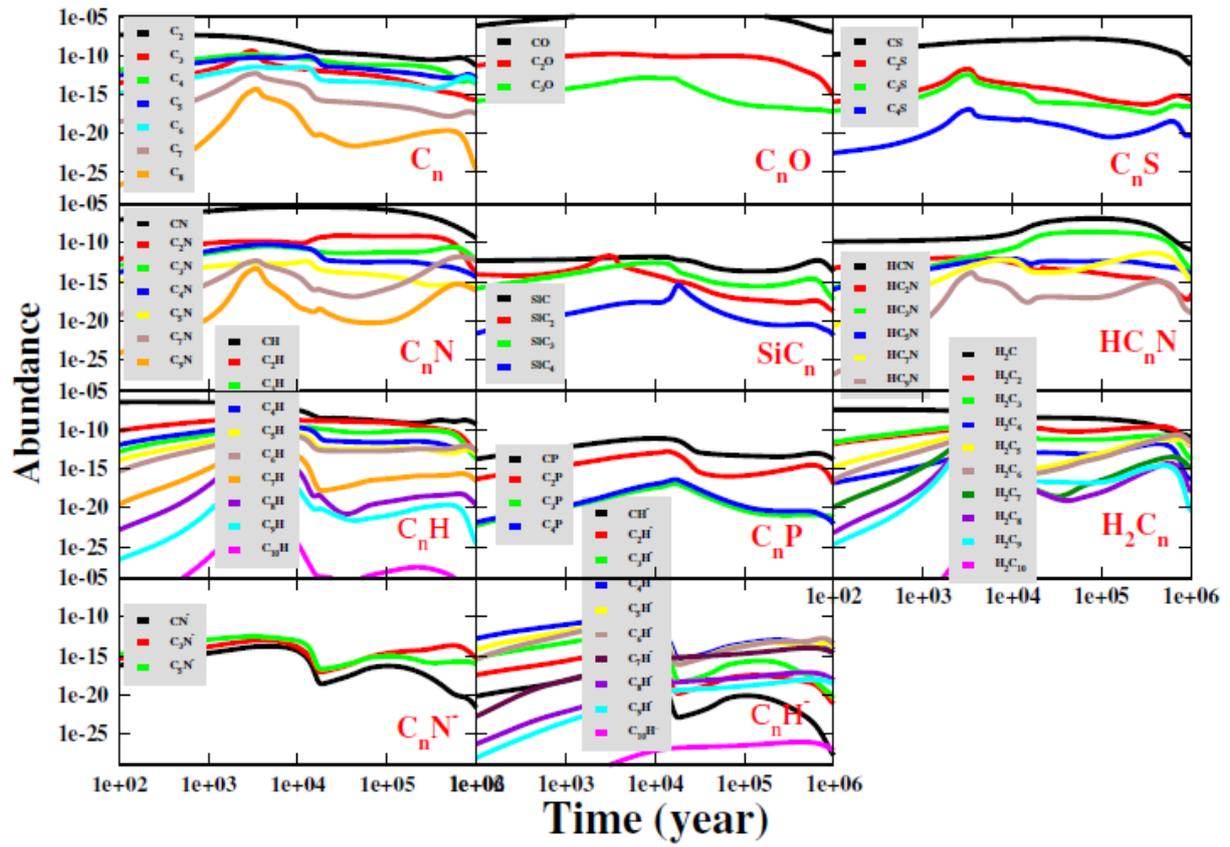

**Figure 15:** Chemical evolution of carbon chains for dense cloud

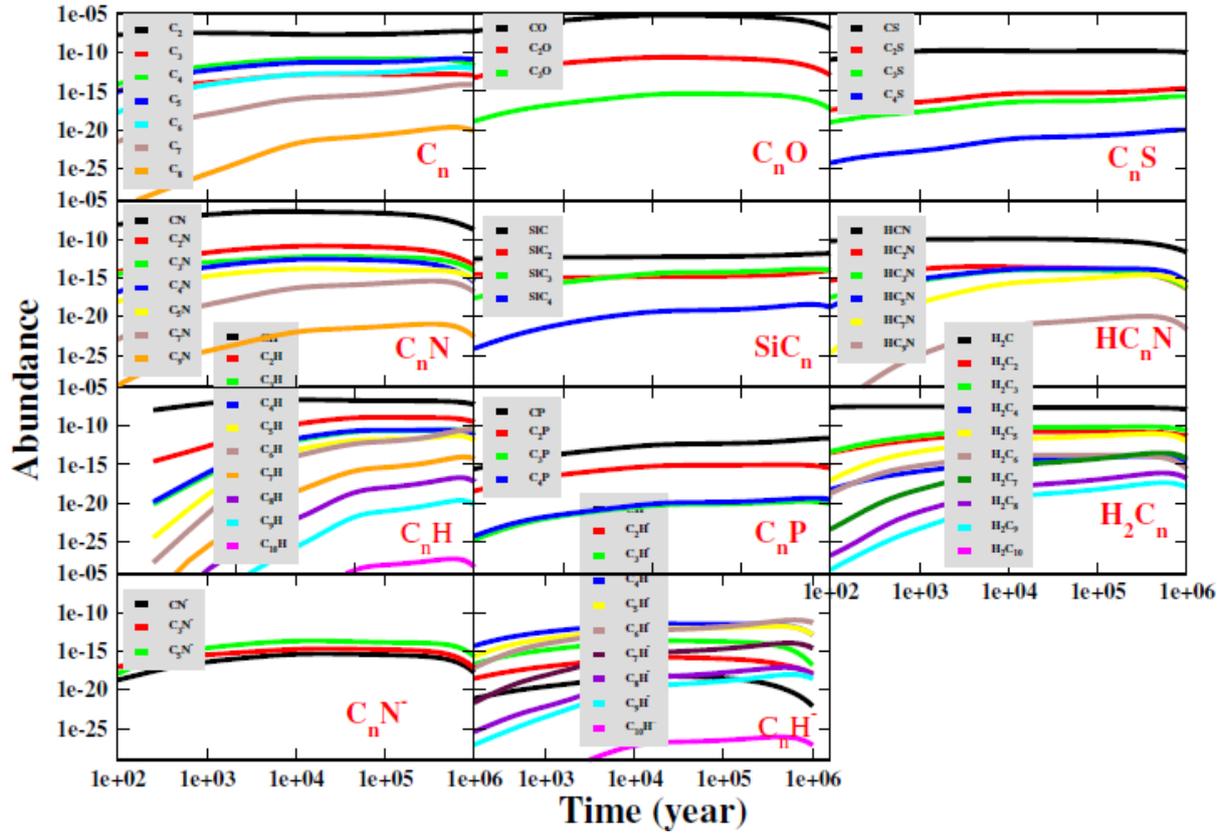

**Figure 16:** Chemical evolution of carbon chains for translucent cloud

Table 15: Abundances of carbon chain molecules.

| Carbon Chains | Peak Abundance (Final abundance) with respect to total H in all forms | |
|---|---|---|
| | **Translucent cloud** | **Dense cloud** |
| $C_2$ | 5.32E-08(4.53E-08) | 4.62E-08 (4.94E-12) |
| $C_3$ | 1.58E-13 (1.17E-13) | 3.97E-10 (3.44E-16) |
| $C_4$ | 1.51E-11 (2.03E-12) | 1.32E-10 (2.31E-14) |
| $C_5$ | 1.69E-11 (1.17E-11) | 8.62E-11 (1.23E-13) |
| $C_6$ | 1.23E-12 (7.90E-13) | 3.69E-12 (4.40E-14) |
| $C_7$ | 7.68E-15 (6.90E-15) | 5.37E-13 (3.66E-18) |
| $C_8$ | 2.40E-20 (7.91E-21) | 4.86E-15 (3.29E-25) |
| | | |
| CO | 6.26E-06 (1.08E-07) | 2.93E-05 (1.09E-07) |
| $C_2O$ | 2.46E-11 (1.23E-13) | 1.67E-10 (7.85E-16) |
| $C_3O$ | 5.07E-16 (5.35E-18) | 1.43E-13 (5.85E-18) |
| | | |
| CH | 2.22E-07 (5.68E-08) | 3.72E-07 (5.43E-10) |
| $C_2H$ | 1.19E-09 (2.98E-10) | 2.89E-09 (1.13E-13) |
| $C_3H$ | 2.88E-11 (1.12E-11) | 2.97E-10 (2.49E-14) |
| $C_4H$ | 2.97E-11 (5.16E-12) | 3.51E-10 (1.86E-13) |
| $C_5H$ | 4.86E-12 (1.45E-12) | 2.79E-11 (9.76E-14) |
| $C_6H$ | 3.49E-11 (3.40E-11) | 1.51E-11 (1.96E-13) |
| $C_7H$ | 9.51E-15 (5.67E-15) | 1.23E-12 (2.73E-17) |
| $C_8H$ | 2.08E-17 (6.61E-18) | 1.04E-12 (2.76E-20) |
| | | |
| CN | 3.60E-07 (1.58E-09) | 4.25E-06 (4.08E-10) |
| $C_2N$ | 1.29E-11 (4.14E-14) | 8.32E-10 (3.19E-14) |
| $C_3N$ | 5.88E-13 (7.75E-15) | 3.45E-11 (5.93E-13) |
| $C_4N$ | 2.68E-13 (3.21E-16) | 5.90E-11 (4.11E-15) |
| $C_5N$ | 1.53E-14 (7.44E-16) | 4.14E-13 (2.63E-16) |
| $C_6N$ | | |
| $C_7N$ | 4.16E-16 (1.62E-17) | 1.61E-12 (4.86E-13) |
| $C_9N$ | 1.19E-21(2.54E-23) | 4.64E-14(7.61E-17) |
| | | |
| CP | 2.53E-12 (2.53E-12) | 7.83E-12 (1.63E-14) |
| $C_2P$ | 9.57E-16 (3.55E-16) | 1.79E-13 (1.21E-17) |
| $C_3P$ | 1.82E-20 (9.47E-21) | 1.31E-17 (6.57E-22) |
| $C_4P$ | 4.83E-20 (3.27E-20) | 4.45E-17 (9.13E-23) |
| $C_5P$ | | |
| $C_6P$ | | |
| | | |
| HCN | 1.06E-10 (2.06E-12) | 1.33E-07 (1.27E-11) |
| $HC_2N$ | 3.06E-14 (4.14E-17) | 1.21E-12 (1.23E-17) |
| $HC_3N$ | 1.13E-14 (5.67E-17) | 2.46E-09 (3.28E-14) |
| $HC_5N$ | 1.66E-14 (2.67E-16) | 8.37E-13 (1.10E-14) |
| $HC_7N$ | 2.37E-15 (1.06E-16) | 4.86E-12 (1.35E-15) |
| $HC_9N$ | 1.17E-20 (2.36E-22) | 1.53E-14 (1.03E-19) |
| | | |
| $H_2C_2$ | 1.76E-11 (4.77E-12) | 2.75E-10 (8.38E-13) |
| $H_2C_3$ | 2.33E-11(7.51E-11) | 4.48E-10 (4.95E-14) |
| $H_2C_4$ | 1.21E-14 (2.87E-15) | 1.69E-12 (3.58E-17) |
| $H_2C_5$ | 8.96E-12 (9.26E-13) | 5.64E-11 (1.05E-12) |
| $H_2C_6$ | 1.90E-11 (7.85E-12) | 2.11E-11 (2.05E-12) |
| $H_2C_7$ | 3.18E-14 (5.86E-15) | 5.77E-12 (1.06E-15) |
| $H_2C_8$ | 8.27E-17 (1.71E-17) | 1.93E-12 (1.22E-18) |
| $H_2C_9$ | 4.73E-18 (1.21E-18) | 1.42E-12(5.32E-21) |
| | | |
| CSi | 1.47E-12 (1.47E-12) | 1.51E-12 (1.97E-14) |
| $C_2Si$ | 1.04E-14 (1.04E-14) | 2.14E-12 (1.64E-19) |
| $C_3Si$ | 1.48E-14 (1.07E-14) | 3.57E-16 (5.17E-18) |
| $C_4Si$ | 4.04E-19 (1.73E-19) | 3.57E-16 (1.32E-22) |
| | | |
| CS | 1.79E-10 (9.97E-11) | 1.58E-08 (7.11E-12) |
| $C_2S$ | 2.32E-15 (2.32E-15) | 1.93E-12 (2.46E-16) |
| $C_3S$ | 2.24E-16 (2.19E-16) | 3.53E-13 (3.98E-17) |
| $C_4S$ | 1.07E-20 (1.04E-20) | 1.27E-17 (5.04E-21) |
| | | |
| $CN^-$ | 4.71E-16 (1.82E-18) | 1.85E-14 (2.52E-22) |
| $C_3N^-$ | 2.03E-15 (6.99E-18) | 1.07E-13 (5.09E-16) |
| $C_5N^-$ | 2.12E-14 (2.01E-16) | 3.46E-13 (1.04E-16) |
| | | |
| $CH^-$ | 6.46E-19 (8.64E-23) | 1.30E-16 (1.80E-28) |
| $C_2H^-$ | 1.65E-16 (1.34E-18) | 1.07E-14 (7.42E-22) |
| $C_3H^-$ | 2.24E-14 (1.49E-17) | 6.53E-13 (4.20E-21) |
| $C_4H^-$ | 3.32E-12 (1.29E-13) | 4.46E-11 (6.74E-15) |
| $C_5H^-$ | 1.58E-12 (1.35E-13) | 1.31E-11 (1.32E-14) |
| $C_6H^-$ | 1.00E-11 (4.81E-12) | 1.02E-11 (4.01E-14) |
| $C_7H^-$ | 1.11E-14 (2.44E-15) | 2.81E-12 (1.71E-17) |
| $C_8H^-$ | 7.99E-18 (8.47E-19) | 7.78E-13 (5.64E-21) |
| $C_9H^-$ | 1.01E-18 (5.61E-23) | 3.05E-13 (2.93E-19) |

**Conclusion**: A comprehensive investigation of different interstellar carbon chain molecules; $C_n$, $H_2C_n$, $HC_nN$ and $C_nX$ (X=N, O, Si, S, H, P, H$^-$, N$^-$) with a total of 89 species have been carried out using high level quantum chemical simulation to determine their accurate enthalpies of formation, chemical reactivity indices; global hardness and softness; and other chemical parameters of these molecules. Chemical modeling of the abundances of these molecular species has also been performed. From the results, in all the groups considered, the most stable molecules have been astronomically observed. The next possible candidate for astronomical observations in each of the carbon chain groups considered is proposed. The effect of kinetics in the formation of some of these molecular species is shown to be correlated with our observations.

**Acknowledgement**: EEE acknowledges a research fellowship from the Indian Institute of Science, Bangalore. AD, PG & SKC are grateful to DST (Grant No. SB/S2/HEP-021/2013) for the partial financial support.